\documentclass[
final
]{elsarticle}
\usepackage[
reviewcopy
]{adndt}

\usepackage{longtable}

\usepackage[utf8]{inputenc}
\usepackage[T1]{fontenc}
\usepackage{microtype}

\usepackage{xcolor}
\usepackage[colorlinks=true,linkcolor=blue,urlcolor=blue,citecolor=blue,bookmarks=true,bookmarksopenlevel=2,breaklinks=true]{hyperref}

\usepackage{mathptmx}

\usepackage{amsmath}
\usepackage{amssymb}
\usepackage{array,booktabs,tabularx}

\biboptions{square,sort&compress}
\bibpunct[]{[}{]}{,}{n}{}{;}
\usepackage{xspace}
\newcommand{\grasp}{\textsc{Grasp2018}\xspace}
\newcommand{\fac}{\textsc{Fac}\xspace}
\newcommand{\mcdfgme}{\textsc{Mcdfgme}\xspace}
\newcommand{\ket}[1]{\left\vert#1\right\rangle}

\usepackage{mhchem}
\usepackage{siunitx}
\usepackage[normalem]{ulem}
\usepackage{soul}
\soulregister\cite7
\soulregister\ref7
\soulregister\pageref7
\usepackage{tikz}
\usepackage{pgfplots}
\pgfplotsset{compat=1.8}
\usepgfplotslibrary{dateplot}
\usepackage{sansmath}

\begin{document}

\begin{frontmatter}

\journal{Atomic Data and Nuclear Data Tables}

%% Author, fill in article title here

\title{Ionization energies of \ce{W^2+} through \ce{W^71+}}

%% Fill in author list here
  \author{Andrzej Brosławski}

  \author{Karol Kozio{\l}\corref{cor}}
  \ead{E-mail: karol.koziol@ncbj.gov.pl}

  \cortext[cor]{Corresponding author.}

  \author{Jacek Rzadkiewicz}

  \address{Narodowe Centrum Bada\'{n} J\k{a}drowych (NCBJ), Andrzeja So{\l}tana 7, 05-400 Otwock-\'{S}wierk, Poland}

\date{2023} %please do not use \today, use actual date of version

\begin{abstract}   
High-accuracy Multi-Configuration Dirac--Hartree--Fock calculations of ionization energies with Configuration Interaction have been carried out for tungsten ions (\ce{W^2+} to \ce{W^71+}) by means of the \grasp code. A comparison with other available experimental and theoretical data, including those calculated by \fac code based on the Dirac--Hartree--Fock--Slater method and with the NIST database has been presented. This work provides a new set of recommended ionization energy values for tungsten ions using state-of-the-art theoretical calculations with uncertainties ranging from 0.07~eV to 0.4~eV, which seems to be a significant improvement over the previous reference uncertainties.
\end{abstract}

\end{frontmatter}

\clearpage

\tableofcontents
\listofDtables
\listofDfigures
\vskip5pc

%%%%%

\section{Introduction}

Tungsten, with its low erosion rate, low tritium retention, high melting point, high thermal resistance and thus high power handling capability has been chosen as a plasma facing component (PFC) for the divertor of ITER \cite{Hirai2013,Aymar2002}. As a result of sputtering at the surfaces with plasma contact, tungsten impurity can appear in the plasma. Since the electron temperatures in ITER will range from tenths of keV to as much as 40~keV in the plasma core \cite{Putterich2008,ITER_research_plan}, a wide set of tungsten ionization degrees will be present. Therefore, for modelling calculations and for diagnostics of key plasma parameters, it is important to know highly accurate atomic data for all stages of ionization of tungsten, with ionization energies (IE) at the top of the list. To date, the most accurate data on the ionization energy of tungsten ions are collected in NIST tables \cite{Kramida2006,NIST_ASD}. Most of the data collected in tables are based on the semi-empirical method.

The most accurate values of IE for tungsten can be obtained for H-like \ce{W^73+} (\SI{80755.91+-0.14}{eV}) and He-like \ce{W^72+} ions (\SI{79181.94+-0.10}{eV}) by means of ab inito quantum electrodynamics calculations \cite{Yerokhin2015,Artemyev2005}. These values correspond to the upmost relative uncertainties of 1.7~ppm for \ce{W^73+} and 1.3~ppm for \ce{W^72+}, respectively (see Table~\ref{tab:compare-ip}). The uncertainties of these theoretical values are mainly due to  the two-loop QED corrections and the lack of detailed knowledge of the nuclear charge distributions. The high accurate data are also available for Li-like \ce{W^71+} IE = \SI{19686.74+-0.20}{eV} (10~ppm relative uncertainty) \cite{Sapirstein2011}. A significant decrease in the accuracy of the NIST reference values of ionization energy is visible from Be-like \ce{W^70+} (\SI{19362+-3}{eV}) to C-like \ce{W^68+} (\SI{18476+-4}{eV}) and a further decrease from N-like \ce{W^67+} (\SI{16588+-11}{eV}) to Ne-like \ce{W^64+} (\SI{15566+-11}{eV}) \cite{NIST_ASD}. The high precision QED theoretical calculations of IE were reported only for Be-like and B-like spaces, namely \SI{19359.25+-0.07}{eV} and for \ce{W^70+} \cite{Malyshev2015} and \SI{18823.28+-0.05}{eV} \ce{W^69+} ions \cite{Malyshev2017}, respectively. Comparison of NIST reference values \cite{Kramida2006,NIST_ASD} with other data in the literature available for states C-like (\ce{W^68+}) to Ne-like (\ce{W^64+}) in Table~\ref{tab:compare-ip} \cite{Malyshev2015,Malyshev2017,Beiersdorfer2012} shows large differences in IEs ranging from $\Delta\text{IE}=\left(\text{IE}_\text{NIST}-\text{IE}_\text{Other}\right)$ = \SI{58}{eV} for C-like \ce{W^68+} to $\Delta\text{IE}$ = \SI{-70}{eV} for F-like \ce{W^65+} ions. This comparison indicates that the reported accuracy of NIST data for some high ionization states of tungsten may be underestimated. It is therefore clear that verification of the IE values for tungsten in this range of charge states is highly desirable. No experimental data on direct measurements of IEs are available in this charge state range. Only a few X-ray transition values are available for He-, Li- and Ne-like tungsten ions (see Table~\ref{tab:compare-tr}). These data allow only for indirect verification of the IE theoretical values.

For \ce{W^63+} to \ce{W^46+} ions having ground state configurations with the principal quantum number $n = 3$, the uncertainties of the IEs presented in the NIST database are \SIrange{3}{7}{eV}, corresponding to a relative uncertainty of $\sim$0.1\%. Comparison with other data available in the literature for this range of ions \cite{Beiersdorfer2012} generally shows discrepancies larger than those reported in NIST. In the case of \ce{W^57+} and \ce{W^54+} ions, these differences reach as much as $\Delta\text{IE}$ = \SI{18}{eV}. For Ni-like \ce{W^46+} one can find the experimental value $\text{IE}_\text{expt} \left( \ce{W^46+} \right)$ = \SI{4057+-3}{eV} derived from measured Rydberg transitions induced by laser irradiation of the tungsten target. For \ce{W^46+} ions there are also the accurate experimental energies of ground-state x-ray transitions measured using high-resolution x-ray spectroscopy at electron-beam ion trap (EBIT) facility \cite{Clementson2010a,Clementson2010b} (see Table~\ref{tab:compare-tr}). These energies can be used for indirect verification of IE values. 

The IEs of tungsten ions having ground state configurations with the principal quantum number $n=4$ (\ce{W^45+} through \ce{W^17+}) are determined in NIST with uncertainties of \SIrange{1.4}{2.1}{eV} (0.06\%-0.33\% relative uncertainties) \cite{Kramida2006,NIST_ASD}. The uncertainty of IE for Cu-like \ce{W^45+} (IE = \SI{2414.1}{eV}) that was determined experimentally is \SI{0.4}{eV} (0.017\% relative uncertainty) \cite{Seely1989a}. In this range of tungsten charge states, the differences between data from NIST and Ref. \cite{Beiersdorfer2012} can be significantly higher than the reference uncertainties. The differences in IE values range from \SI{-4.1}{eV} for Br-like \ce{W^39+} to \SI{+9.4}{eV} for Sr-like \ce{W^36+}. 

The uncertainties of the IEs for lowest charge states (\ce{W^16+} through \ce{W^2+}) range from \SI{0.3}{eV} for \ce{W^4+} to \SI{1.5}{eV} for \ce{W^{14+,15+}}. These values correspond to 0.3\% to 1.5\% relative uncertainties. The experimentally determined uncertainties of IEs for Er-like (\ce{W^6+}) and Tm-like (\ce{W^5+}) are \SI{0.06}{eV} and \SI{0.04}{eV} \cite{Sugar1975,Sugar1979}. However, these ionization energies were determined by applying the Rydberg-Ritz series formula, where the uncertainty is attributed to $4f^{13}ns$ series. Therefore, one can expect the verification of the reported uncertainties.     

Here we present the high-accuracy MCDHF-CI calculations of the IEs for  \ce{W^2+} to \ce{W^71+} tungsten ions by means the \grasp code \cite{FroeseFischer2018}. The influence of electron correlation on the W ionization energies is taken into account in the manner presented in our previous paper \cite{Kozio2021}. The presented values have been compared with the previous reference IE values  provided by Kramida and Reader \cite{Kramida2006,NIST_ASD} and with other available theoretical and experimental values, in particular with values provided by Beiersdorfer et al. \cite{Beiersdorfer2012}. In addition, new IE values were also compared with those obtained with the \fac code calculations \cite{Gu2008}. We believe, that new data with uncertainties up to two orders of magnitude lower than the previous reference values \cite{Kramida2006,NIST_ASD} should be a complete set of recommended IE values useful for many physics applications.

\begin{table}[!htb]
\caption{Collection of theoretical (theo.), semi-empirical (s-emp.), and experimental (exp.) results of IE for selected W ions (in eV) found in the literature.}
\label{tab:compare-ip}
\begin{tabular*}{\linewidth}{@{}l @{\extracolsep{\fill}} l ll@{}}
\toprule
Ion & Sequence & Ref. \cite{Kramida2006} & Other \\
\midrule
\ce{W^71+} & Li & 19691$\pm$3 & 19686.74$\pm$0.2 \cite{Sapirstein2011} (theo.), 19686.94 \cite{Beiersdorfer2012} (theo.) \\
\ce{W^70+} & Be & 19362$\pm$3 & 19359.25$\pm$0.07 \cite{Malyshev2015} (theo.), 19348.69 \cite{Beiersdorfer2012} (theo.) \\
\ce{W^69+} & B & 18872$\pm$3 & 18823.28$\pm$0.05 \cite{Malyshev2017} (theo.), 18829.80 \cite{Beiersdorfer2012} (theo.) \\
\ce{W^68+} & C & 18476$\pm$4 & 18417.59 \cite{Beiersdorfer2012} (theo.) \\
\ce{W^67+} & N & 16588$\pm$11 & 16635.31 \cite{Beiersdorfer2012} (theo.) \\
\ce{W^66+} & O & 16252$\pm$11 & 16266.94 \cite{Beiersdorfer2012} (theo.) \\
\ce{W^65+} & F & 15896$\pm$11 & 15965.53 \cite{Beiersdorfer2012} (theo.) \\
\ce{W^64+} & Ne & 15566$\pm$11 & 15603.58 \cite{Beiersdorfer2012} (theo.) \\
\ce{W^63+} & Na & 7130$\pm$7 & 7131.82 \cite{Sapirstein2015} (theo.), 7131.33 \cite{Beiersdorfer2012} (theo.) \\
\ce{W^45+} & Cu & 2414.1$\pm$0.4 & 2414.1$\pm$0.4 \cite{Seely1989a} (s-emp.), 2414.91 \cite{Safronova2012} (theo.) \\
 &  &  & 2412.7$\pm$1.8 \cite{Tragin1989} (exp.), 2413.65 \cite{Tragin1989} (theo.)\\
 &  &  & 2413.51 \cite{Beiersdorfer2012} (theo.) \\
\ce{W^46+} & Ni & 4057$\pm$3 & 4056.8$\pm$3 \cite{Tragin1989} (exp.), 4118.04 \cite{Tragin1989} (theo.) \\
 &  &  & 4051.67 \cite{Beiersdorfer2012} (theo.)\\
\ce{W^37+} & Rb & 1621.7$\pm$1.5 & 1619.77 \cite{Safronova2016} (theo.), 1617.82 \cite{Beiersdorfer2012} (theo.) \\
\ce{W^27+} & Ag & 881.4$\pm$1.6 & 880.96 \cite{Safronova2010} (theo.), 879.19 \cite{Beiersdorfer2012} (theo.) \\
\ce{W^6+} & Er & 122.01$\pm$0.06 & 122.01$\pm$0.06 \cite{Sugar1975} (exp.), 120.55 \cite{Beiersdorfer2012} (theo.) \\
\ce{W^5+} & Tm & 64.77$\pm$0.04 & 64.77$\pm$0.04 \cite{Sugar1979} (exp.), 63.59 \cite{Beiersdorfer2012} (theo.), 66.1 \cite{Muller2019a} (exp.) \\
\bottomrule
\end{tabular*}
\end{table}

\section{Theoretical background}

The calculations of the energy levels for tungsten ions with $q$ = 2--71 have been carried out by means of the \grasp code \cite{FroeseFischer2018}, implementing the Multi-Configuration Dirac--Hartree--Fock (MCDHF) method with Configuration Interaction. 
The methodology of MCDHF calculations performed in the present study is similar to that published earlier, in many papers (see, e.g., \cite{Grant2007}). 

The accuracy of the wavefunction depends on the configuration state functions (CSFs) included in its expansion \cite{FroeseFischer2016}. The accuracy can be improved by extending the CSF set by including the CSFs originating from excitations from orbitals occupied in the reference CSFs to unfilled orbitals of the active orbital set (i.e., CSFs for virtual excited states). This approach is called Configuration Interaction, CI. The CI method makes it possible to include the major part of the electron correlation contribution to the energy of the atomic levels. In the CI approach, it is important to choose an appropriate basis of CSFs for the virtual excited states. It can be done with systematically building CSF sequences by extending the Active Space (AS) of orbitals and concurrently monitoring the convergence of the self-consistent calculations. 

For comparative purposes we performed calculations by using the \fac code \cite{Gu2008}, based on the multiconfigurational Dirac--Hartree--Fock--Slater (MCDHFS) method. In general, the MCDHFS method is similar to the MCDHF one, in that they both refer to the effective Hamiltonian and multiconfigurational ASF. The main difference between the MCDHF and the MCDHFS methods is the approximation of the non-local Dirac--Hartree--Fock exchange potential by a local potential used in the latter. \fac uses an improved form of the local exchange potential (see \cite{Gu2008} for details).

\subsection{Computational details}

In present calculations we have used the active spaces of virtual orbitals with $n$ up to $n$ = 7 (for IEs of \ce{W^2+}--\ce{W^28+}) or up to $n$ = 6 (for the other IEs) and $l$ up to $l$ = 4. 
The excitations patterns for ions are presented in Table~\ref{tab:exc-ci}. It shows atomic orbitals divided into three groups: inactive orbitals, excluded from excitations (labeled I), fully active orbitals, from which single and double excitations are allowed (labeled SD), and partially active orbitals, from which only single excitations are allowed (labeled S).
The ionization energy has been calculated as a difference between the energies of the lowest levels of ions $W^q$ and $W^{q+1}$, related to their groundstate electronic configurations. 
The groundstate configurations are also listed in Table~\ref{tab:exc-ci}. 

Because for the IEs relating to the removing of the electron from the most inner shells the QED effects are very important in precise evaluating of such values, for calculating of IE for \ce{W^71+} and \ce{W^70+} we also used higher order QED corrections, calculated by means of \mcdfgme code \cite{mcdfgme}, that are omitted in the \grasp code calculations. They are Wichmann-Kroll vacuum polarization term and two-loop self-energy and mixed self-energy and vacuum polarization corrections, as well as recoil correction. For \ce{W^71+} and \ce{W^70+} they contribute $\sim$\SI{0.14}{eV} to IE values, but for ions with $q<70$ they contribute $\le0.01$ eV. 

The frequency-dependent Breit term contributions have been calculated for all ions at MR level and then added to the final CI values. 
Usually this contribution is very small, but it is significant in the cases when IE is linked to the removing of inner-shell $p$-type electron \cite{Kozio2020}. For IEs of \ce{W^64+}--\ce{W^67+} the frequency-dependent Breit term contribute \SIrange{1.86}{2.02}{eV} and for IEs of \ce{W^56+}--\ce{W^59+} it contribute \SIrange{0.44}{0.48}{eV}. 

\subsection{Examination of MCDHF-CI calculations}

\begin{figure}
\centering
\begin{tikzpicture}[
% x=1cm,y=0.01pt,xscale=1.0,yscale=1.0, 
font=\sffamily\sansmath]
\begin{axis}[%height=6cm,width=10cm,
xlabel={Active Space},x label style={font=\large\sffamily\sansmath},
ylabel={E (eV)},y label style={font=\large\sffamily\sansmath},
xmin=-0.2,xmax=4.2,
xtick={0,1,2,3,4},
xticklabels={MR,AS1,AS2,AS3,AS4},
y tick label style={/pgf/number format/.cd,fixed,fixed zerofill,precision=0,1000 sep={}},
scaled y ticks = false,
legend pos=north east,
legend cell align=left
%ymajorgrids=true,grid style=dashed
]
 \addplot[color=red,mark=o] coordinates 
 {
 (0,1699.3999)
 (1,1696.2835)
 (2,1696.2251)
 (3,1696.2176)
 (4,1696.2125)
 };
\addlegendentry{\makebox{$\ket{1s^2 2p_{3/2}}_{3/2} \to \ket{1s^2 2s_{1/2}}_{1/2}$}};
\end{axis}
\end{tikzpicture}
\hfill
\begin{tikzpicture}[
% x=1cm,y=0.01pt,xscale=1.0,yscale=1.0, 
font=\sffamily\sansmath]
\begin{axis}[%height=6cm,width=10cm,
xlabel={Active Space},x label style={font=\large\sffamily\sansmath},
ylabel={E (eV)},y label style={font=\large\sffamily\sansmath},
xmin=-0.2,xmax=4.2,
xtick={0,1,2,3,4},
xticklabels={MR,AS1,AS2,AS3,AS4},
y tick label style={/pgf/number format/.cd,fixed,fixed zerofill,precision=0,1000 sep={}},
scaled y ticks = false,
legend pos=south east,
legend cell align=left
%ymajorgrids=true,grid style=dashed
]
 \addplot[color=red,mark=o] coordinates 
 {
 (0,1732.2003)
 (1,1743.0837)
 (2,1741.6471)
 (3,1741.6311)
 (4,1741.5662)
 };
\addlegendentry{\makebox{$\ket{1s^2 2s_{1/2} 2p_{3/2}}_{1} \to \ket{1s^2 2s^2}_{0}$}};
\end{axis}
\end{tikzpicture}
\caption{Convergence of CI calculations for $\ket{1s^2 2p_{3/2}}_{3/2} \to \ket{1s^2 2s_{1/2}}_{1/2}$ transition in Li-like W (left) and $\ket{1s^2 2s_{1/2} 2p_{3/2}}_{1} \to \ket{1s^2 2s^2}_{0}$ transition in Be-like W (right).}\label{fig:tr_libe_conv}
\end{figure}
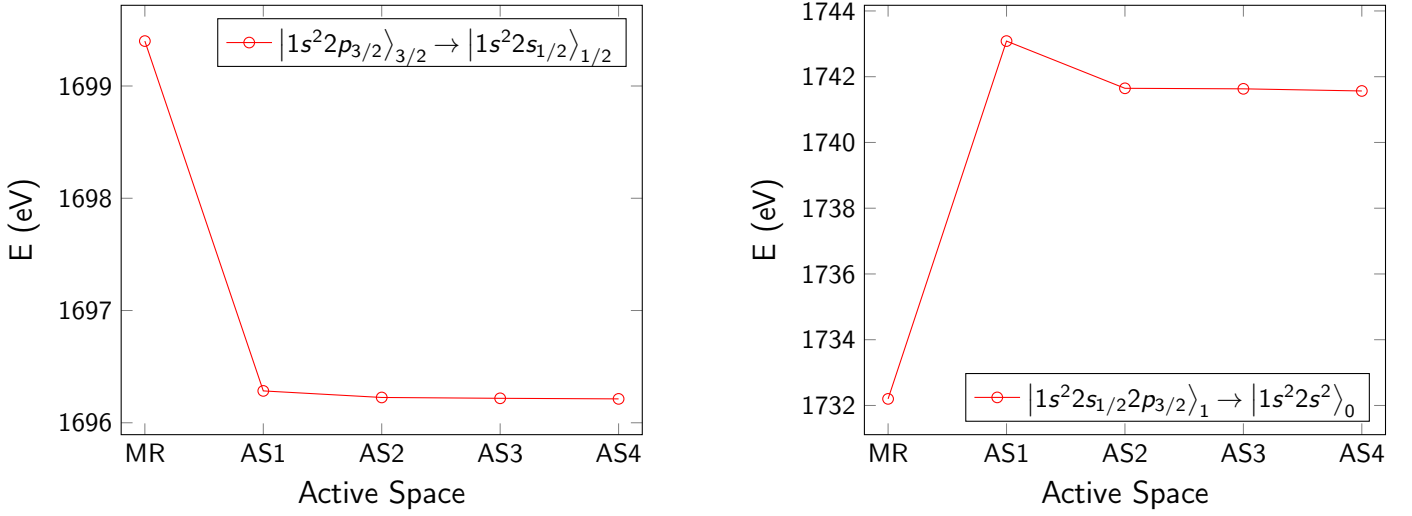

In order to examine our MCDHF-CI calculations we compared the theoretical predictions of the ground-state transition energies with the measured ones in Li-, Be, and Ne-like W ions (see Table~\ref{tab:compare-tr}). Fig.~\ref{fig:tr_libe_conv} shows the convergence of MCDF-CI calculations for $\ket{1s^2 2p_{3/2}}_{3/2} \to \ket{1s^2 2s_{1/2}}_{1/2}$ and $\ket{1s^2 2s_{1/2} 2p_{3/2}}_{1} \to \ket{1s^2 2s^2}_{0}$ transitions in Li-like and Be-like W ions. In both cases, one can see the convergence from the AS2 active space. The final values (AS4) of the transition energies in Li-like and Be-like W ions, namely \SI{1696.21}{eV} and \SI{1741.57}{eV} are consistent within the limits of experimental uncertainties \SIrange{0.5}{1.3}{eV} (see Table~\ref{tab:compare-tr}). The same is true for energies of other considered transitions in \ce{W^64+} and \ce{W^46+} ions. Our MCDF-CI calculations are also consistent with other accurate Dirac--Fock calculations. The calculations presented in Refs. \cite{Ivanova1991,Aggarwal2016,Dong2003} seem to be slightly less accurate. In addition, we compared the \fac calculations with the same experimental transition energies. For these calculations slightly lower accuracy (\SI{+-3}{eV}) can be seen for all considered transitions. Therefore, one can conclude that the MCDHF-CI calculations provide the most accurate predictions of the level energies, including the IEs.

\begin{table}[!htb]
\caption{Comparison of present MCDHF-CI calculation results for selected transitions in W ions (in eV) to the data found in the literature.}
\label{tab:compare-tr}
\begin{tabular*}{\linewidth}{@{}l @{\extracolsep{\fill}} l llll@{}}
\toprule
& & \multicolumn{2}{c}{Present} & \multicolumn{2}{c}{Literature}\\
\cmidrule{3-4}\cmidrule{5-6}
Ion & Transition & MCDHF-CI & FAC-CI & Experimental & Other theoretical\\
\midrule
\ce{W^71+} & $\ket{1s^2 2p_{3/2}}_{3/2} \to \ket{1s^2 2s_{1/2}}_{1/2}$ & 1696.21 & 1698.43* & 1696.2(5) \cite{Clementson2011},  & 1695.9946 \cite{Kim1991}, 1696.0309 \cite{Chen1995}, \\
& & & & 1697(1) \cite{Podpaly2009} & 1695.42 \cite{Safronova2010a} \\
\ce{W^70+} & $\ket{1s^2 2s_{1/2} 2p_{3/2}}_{1} \to \ket{1s^2 2s^2}_{0}$ & 1741.57 & 1744.53* & 1741.4(6) \cite{Clementson2011},  & 1740.538 \cite{Safronova2010}, 1741.47 \cite{Chen1997}, \\
& & & & 1741.08(125) \cite{Podpaly2009} & 1741.371 \cite{Safronova1996}, 1743.179 \cite{Singh2018} \\
\ce{W^64+} & $\ket{1s^2 2s^2 2p_{1/2}^2 2p_{3/2}^3 3s_{1/2}}_{1} \to \ket{1s^2 2s^2 2p^6}_{0}$ & 8307.20 & 8305.15* & 8307.51(40) \cite{Beiersdorfer2012a}  & 8306.76 \cite{Vilkas2008}, 8312.05 \cite{Ivanova1991}, \\
& & & & & 8302.591 \cite{Aggarwal2016} \\
\ce{W^64+} & $\ket{1s^2 2s^2 2p_{1/2}^2 2p_{3/2}^3 3s_{1/2}}_{2} \to \ket{1s^2 2s^2 2p^6}_{0}$ & 8298.75 & 8296.51* & 8299.22(40) \cite{Beiersdorfer2012a}  & 8298.33 \cite{Vilkas2008}, 8303.48 \cite{Ivanova1991}, \\
& & & & & 8294.063 \cite{Aggarwal2016} \\
\ce{W^46+} & $\ket{[Ar] 3d_{3/2}^{4} 3d_{5/2}^{5} 4s_{1/2}}_{3} \to \ket{[Ar] 3d^{10}}_{0}$ & 1562.19 & 1561.46 & 1562.0(3) \cite{Clementson2010b} & 1562.2 \cite{Clementson2010b}, 1560.31 \cite{Ballance2006} \\
& $\ket{[Ar] 3d_{3/2}^{4} 3d_{5/2}^{5} 4s_{1/2}}_{2} \to \ket{[Ar] 3d^{10}}_{0}$ & 1564.04 & 1563.31 & 1563.9(3) \cite{Clementson2010a} & 1564.1 \cite{Clementson2010b}, 1562.14 \cite{Ballance2006}, \\
& & & & & 1561.11 \cite{Dong2003}, 1563.63 \cite{Safronova2006}\\
& $\ket{[Ar] 3d_{3/2}^{3} 3d_{5/2}^{6} 4s_{1/2}}_{2} \to \ket{[Ar] 3d^{10}}_{0}$ & 1629.84 & 1629.18 & 1629.8(3) \cite{Clementson2010a} & 1630.0 \cite{Clementson2010b}, 1629.28 \cite{Ballance2006}, \\
& & & & & 1626.77 \cite{Dong2003}, 1629.62 \cite{Safronova2006}\\
\bottomrule
\end{tabular*}
* Small-base CI with only excitations within $n \le 3$ shells allowed. 
\end{table}

\subsection{Computational strategies to CI calculations for open-shell ions}

Because for tungsten ions having many open shells the number of CSFs increases drastically, there is a need to choose a proper computational technique that provides reliable results and is not so much time- and resource-consuming. 
We examine the effect of electron correlation on the tungsten ionization energies by using IE of \ce{W^5+} as a case study. 
Ionization energy of \ce{W^5+} has been calculated as a difference between energies of the groundstates of \ce{W^5+} ($[Kr]\;4d^{10}\;4f^{14}\;5s^{2}\;5p^{6}\;5d^{1}\ \ {}^{2}D_{3/2}$) and \ce{W^6+} ($[Kr]\;4d^{10}\;4f^{14}\;5s^{2}\;5p^{6}\ \ {}^{1}S_{0}$). 
Three computational strategies to CI calculations have been tested: 
(1) FCI -- Full CI; all corrections to the wavefunction due to excitations from selected occupied orbitals to the active space of virtual orbitals are included in CI procedure. 
(2) CV -- Like FCI, but occupied orbitals are divided into three groups: valence (V), active core (C), and inactive core (I). Excitations from C and V orbitals may be differentiated by type. Excitations from I orbitals are prohibited. 
(3) ZF -- CI in which the active space of virtual orbitals is divided into two groups: zero- and first-order spaces. 
The zero-order space contains configuration state functions (CSFs) that account for the major parts of the wavefunction. The first-order space contains CSFs that represent minor corrections. These minor corrections are calculated by applying second-order perturbation theory. This kind of perturbative techniques (i.e. zero- and first-order partitions) is suggested in the cases with a large number of CSFs in MCDHF-CI calculations \cite{Gustafsson2017}. 

\begin{figure}
\centering
\begin{tikzpicture}[scale=1.0,font=\sffamily\sansmath]
\begin{axis}[
xlabel={Active Space},x label style={font=\large\sffamily\sansmath},
ylabel={IE (eV)},y label style={font=\large\sffamily\sansmath},
xmin=-0.2,xmax=4.2,
xtick={0,1,2,3,4},
xticklabels={MR,AS1,AS2,AS3,AS4},
ymin=62,ymax=65,
y tick label style={/pgf/number format/.cd,fixed,fixed zerofill,precision=1},
legend pos=south east,
legend cell align=left
%ymajorgrids=true,grid style=dashed
]

\addplot[color=red,mark=o] coordinates 
{
(0,63.60)
(1,62.38)
(2,63.83)
(3,64.21)
(4,64.31)
};
\addlegendentry{FCI};

\addplot[color=blue,mark=square] coordinates 
{
(0,63.60)
(1,62.76)
(2,64.12)
(3,64.38)
(4,64.45)
};
\addlegendentry{CV};

\addplot[color=green!50!black,mark=triangle] coordinates 
{
(0,63.60)
(1,62.38)
(2,64.20)
(3,64.69)
(4,64.87)
};
\addlegendentry{ZF};

\end{axis}
\end{tikzpicture}
\caption{Computational strategies to CI calculations for IE of \ce{W^5+}.}\label{fig:ci_strategies}
\end{figure}
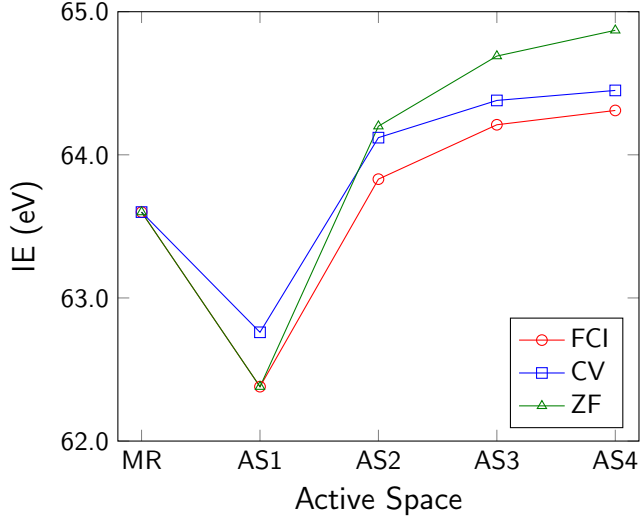

Table~\ref{tab:ci_strategies} collects the ASs used in tested computational strategies. The results of three kind of CI calculations are presented on Fig.~\ref{fig:ci_strategies}. As one can see from this figure, the ZF approach results diverge from the FCI result for higher ASs, but the CV results are still close to the FCI ones. Therefore we conclude that the CV approach can be used when the FCI approach is too time- and resource-consuming. Thus we used CV approach in our calculations of IEs of tungsten ions. 

\begin{table}[!htb]
\caption{Computational strategies to CI calculations for IE of \ce{W^5+}.}
\label{tab:ci_strategies}
\begin{tabular*}{\linewidth}{@{}l @{\extracolsep{\fill}} llll@{}}
\toprule
Model & \multicolumn{4}{c}{Active Space} \\
\cmidrule{2-5}
& \multicolumn{2}{c}{occupied orbitals} & \multicolumn{2}{c}{virtual orbitals} \\
\cmidrule{2-3}\cmidrule{4-5}
& SD exc. & S only exc. & zero-order & first-order \\
\midrule
FCI & $4f$, $5s$, $5p$, $5d$ &  & $n\le8$, $l\le g$ &  \\
CV & $5p$, $5d$ & $4f$, $5s$ & $n\le8$, $l\le g$ &  \\
ZF & $4f$, $5s$, $5p$, $5d$ &  & $n=5$, $l\le g$ & $n=6{-}8$, $l\le g$ \\
\bottomrule
\end{tabular*}
\end{table}

\subsection{Estimating theoretical uncertainties}

The theoretical uncertainties of IEs related to convergence with the size of a basis set have been estimated from theoretical uncertainties of the energy of appropriate atomic levels for given ions. These uncertainties (labeled $\delta E_1$ and $\delta E_2$) have been estimated as absolute value of difference between energies calculated within converged values ($AS\infty$; asymptote value assuming that correlation effects on energy levels are saturated, i.e. $|E(ASn{+}1)-E(ASn)|\to0$ when $n\to\infty$) and the highest ASn for given ions, i.e. $\delta E = |E^{AS\infty}-E^{ASn}|$. 
Such an uncertainty does not exceed \SI{0.4}{eV} in all cases, being even smaller ($\le0.1$ eV) for higher ionized ions having only a few electrons. 
For the IE of \ce{W^5+} the difference between values obtained from FCI and CV models is \SI{0.14}{eV}. This value, $\delta_{FCI{-}CV}$, is added to the uncertainties for ions for which the CV approach has been used, i.e. for \ce{W^2+} through \ce{W^63+}. 
If IE is related to the ionizing of the inner-shell electron ($2s$ or $2p$) the quality of calculation of the radial wavefunction may produce the uncertainty of similar size or even larger than the uncertainty related to the electron correlation effect. In order to estimate the radial wavefunction quality uncertainty, $\delta_{RWF}$, we compared the IE calculated by two MCDHF codes, the \grasp code and the \mcdfgme code \cite{mcdfgme}, without involving extensive CI calculations. Such an uncertainty does not exceed \SI{0.08}{eV} in all cases, being negligible in the cases of IE of low-ionized W ions. 
Finally, we estimate the total theoretical uncertainty of IEs of W ions according to the formula 
\begin{equation*}
\delta(IE) = \sqrt{ (\delta E_1)^2 + (\delta E_2)^2 + (\delta_{FCI{-}CV})^2 + (\delta_{RWF})^2} \;.
\end{equation*}

\section{Recommended IEs and conclusions}

A complete set of recommended IE values for \ce{W^2+} through \ce{W^71+} determined by means of MCDF-CI calculations is presented in Table~\ref{tab:ip}. We also show the reference data collected in the NIST database to date. In addition, the results of the IEs obtained by us from \fac code and those obtained by Beiersdorfer et al. from the relativistic atomic structure calculations (RAC) \cite{Beiersdorfer2012} are also presented. Based on the results presented in this work, a few main conclusions can be drawn. Applying GRASP+CI approach the most accurate IE values with uncertainties ranging from \SI{0.07}{eV} to \SI{0.4}{eV} can be provide, which is a significant improvement over the previous reference uncertainties of up to \SI{11}{eV} \cite{Kramida2006,NIST_ASD}. Our new recommended data agree within the limits of uncertainties with the few available experimental data. It is worth mentioning here the important need for new measurements of IE values for tungsten ions. As for the data obtained by the \fac code, they seem to be consistent with those presented earlier in Ref. \cite{Beiersdorfer2012} and seem to be sufficiently accurate for predictions of IE with an accuracy of several eV.

%\bibliographystyle{apsrev4-1}
%\bibliography{references}

%merlin.mbs apsrev4-1.bst 2010-07-25 4.21a (PWD, AO, DPC) hacked
%Control: key (0)
%Control: author (72) initials jnrlst
%Control: editor formatted (1) identically to author
%Control: production of article title (-1) disabled
%Control: page (0) single
%Control: year (1) truncated
%Control: production of eprint (0) enabled
%

\clearpage

\TableExplanation

\section*{Table 1.}

\begin{tabular}{@{}p{0.5in}p{6in}@{}}
I & inactive orbital\\ 
SD & single and double excitations allowed from given orbital\\
S & only single excitations allowed from given orbital\\
\end{tabular}

\section*{Table 2.}

\begin{tabular}{@{}p{0.7in}p{6in}@{}}
FAC & \fac calculations by means of the Multi-Configuration Dirac--Hartree--Fock--Slater method\\
GRASP+CI & \grasp  calculations by means of the Multi-Configuration Dirac--Hartree--Fock with Configuration Interaction method\\
\end{tabular}

\datatables % This command is necessary to get the table names in toc

\setlength{\LTcapwidth}{\linewidth}
\setlength{\LTleft}{0pt}
\setlength{\LTright}{0pt} 
\setlength{\tabcolsep}{0.5\tabcolsep}
\renewcommand{\arraystretch}{1.0}

\begin{longtable}{@{}l @{\extracolsep{\fill}} l ccccccccccccc@{}}
\caption{Patterns of excitations from occupied orbitals used in CV model of CI calculations.}
\label{tab:exc-ci}
Ion & Groundstate term & \multicolumn{13}{c}{Occupied orbitals} \\
\cmidrule{3-15}
 & & 1s & 2s & 2p & 3s & 3p & 3d & 4s & 4p & 4d & 4f & 5s & 5p & 5d \\
\midrule
\endfirsthead
\multicolumn{15}{l}{Table \ref{tab:exc-ci} (continued)}\\
\midrule
Ion & Groundstate term & \multicolumn{13}{c}{Occupied orbitals} \\
\cmidrule{3-15}
 & & 1s & 2s & 2p & 3s & 3p & 3d & 4s & 4p & 4d & 4f & 5s & 5p & 5d \\
\midrule
\endhead
\midrule
\endfoot
\bottomrule
\endlastfoot
2+ & $[Kr]\;4d^{10}\;4f^{14}\;5s^{2}\;5p^{6}\;5d^{4}\ \ {}^{5}D_{0}$ & I & I & I & I & I & I & I & I & I & S & I & S & SD   \\
3+ & $[Kr]\;4d^{10}\;4f^{14}\;5s^{2}\;5p^{6}\;5d^{3}\ \ {}^{4}F_{3/2}$ & I & I & I & I & I & I & I & I & I & S & I & S & SD   \\
4+ & $[Kr]\;4d^{10}\;4f^{14}\;5s^{2}\;5p^{6}\;5d^{2}\ \ {}^{3}F_{2}$ & I & I & I & I & I & I & I & I & I & S & I & S & SD  \\
5+ & $[Kr]\;4d^{10}\;4f^{14}\;5s^{2}\;5p^{6}\;5d^{1}\ \ {}^{2}D_{3/2}$ & I & I & I & I & I & I & I & I & I & S & I & S & SD  \\
 &  & I & I & I & I & I & I & I & I & I & S & S & SD & SD  \\
6+ & $[Kr]\;4d^{10}\;4f^{14}\;5s^{2}\;5p^{6}\ \ {}^{1}S_{0}$ & I & I & I & I & I & I & I & I & I & S & S & SD &  \\
 &  & I & I & I & I & I & I & I & I & I & SD & S & SD &  \\
7+ & $[Kr]\;4d^{10}\;4f^{13}\;5s^{2}\;5p^{6}\ \ {}^{2}F_{7/2}$ & I & I & I & I & I & I & I & I & I & SD & S & SD &  \\
8+ & $[Kr]\;4d^{10}\;4f^{14}\;5s^{2}\;5p^{4}\ \ {}^{3}P_{2}$ & I & I & I & I & I & I & I & I & I & SD & S & SD &  \\
9+ & $[Kr]\;4d^{10}\;4f^{14}\;5s^{2}\;5p^{3}\ \ {}^{2}P_{3/2}$ & I & I & I & I & I & I & I & I & I & SD & S & SD &  \\
10+ & $[Kr]\;4d^{10}\;4f^{14}\;5s^{2}\;5p^{2}\ \ {}^{3}P_{0}$ & I & I & I & I & I & I & I & I & I & SD & S & SD &  \\
11+ & $[Kr]\;4d^{10}\;4f^{13}\;5s^{2}\;5p^{2}\ \ {}^{4}F_{7/2}$ & I & I & I & I & I & I & I & I & I & SD & S & SD &  \\
 &  & I & I & I & I & I & I & I & I & I & SD & SD & SD &  \\
12+ & $[Kr]\;4d^{10}\;4f^{14}\;5s^{2}\ \ {}^{1}S_{0}$ & I & I & I & I & I & I & I & I & I & SD & SD &  &  \\
13+ & $[Kr]\;4d^{10}\;4f^{13}\;5s^{2}\ \ {}^{2}F_{7/2}$ & I & I & I & I & I & I & I & I & I & SD & SD &  &  \\
14+ & $[Kr]\;4d^{10}\;4f^{12}\;5s^{2}\ \ {}^{3}H_{6}$ & I & I & I & I & I & I & I & I & I & SD & SD &  &  \\
15+ & $[Kr]\;4d^{10}\;4f^{11}\;5s^{2}\ \ {}^{4}I_{15/2}$ & I & I & I & I & I & I & I & I & I & SD & SD &  &  \\
16+ & $[Kr]\;4d^{10}\;4f^{11}\;5s^{1}\ \ {}^{5}I_{8}$ & I & I & I & I & I & I & I & I & I & SD & SD &  &  \\
 &  & I & I & I & I & I & I & I & I & S & SD & SD &  &  \\
17+ & $[Kr]\;4d^{10}\;4f^{11}\ \ {}^{4}I_{15/2}$ & I & I & I & I & I & I & I & I & S & SD &  &  &  \\
18+ & $[Kr]\;4d^{10}\;4f^{10}\ \ {}^{5}I_{8}$ & I & I & I & I & I & I & I & I & S & SD &  &  &  \\
19+ & $[Kr]\;4d^{10}\;4f^{9}\ \ {}^{6}H_{15/2}$ & I & I & I & I & I & I & I & I & S & SD &  &  &  \\
20+ & $[Kr]\;4d^{10}\;4f^{8}\ \ {}^{7}F_{6}$ & I & I & I & I & I & I & I & I & S & SD &  &  &  \\
21+ & $[Kr]\;4d^{10}\;4f^{7}\ \ {}^{8}S_{7/2}$ & I & I & I & I & I & I & I & I & S & SD &  &  &  \\
22+ & $[Kr]\;4d^{10}\;4f^{6}\ \ {}^{7}F_{0}$ & I & I & I & I & I & I & I & I & S & SD &  &  &  \\
23+ & $[Kr]\;4d^{10}\;4f^{5}\ \ {}^{6}H_{5/2}$ & I & I & I & I & I & I & I & I & S & SD &  &  &  \\
24+ & $[Kr]\;4d^{10}\;4f^{4}\ \ {}^{5}I_{4}$ & I & I & I & I & I & I & I & I & S & SD &  &  &  \\
25+ & $[Kr]\;4d^{10}\;4f^{3}\ \ {}^{4}I_{9/2}$ & I & I & I & I & I & I & I & I & S & SD &  &  &  \\
26+ & $[Kr]\;4d^{10}\;4f^{2}\ \ {}^{3}H_{4}$ & I & I & I & I & I & I & I & I & S & SD &  &  &  \\
 &  & I & I & I & I & I & I & I & I & SD & SD &  &  &  \\
27+ & $[Kr]\;4d^{10}\;4f^{1}\ \ {}^{2}F_{5/2}$ & I & I & I & I & I & I & I & I & SD & SD &  &  &  \\
28+ & $[Kr]\;4d^{10}\ \ {}^{1}S_{0}$ & I & I & I & I & I & I & I & I & SD &  &  &  &  \\
 &  & I & I & I & I & I & I & I & SD & SD &  &  &  &  \\
29+ & $[Ar]\;3d^{10}\;4s^{2}\;4p^{6}\;4d^{9}\;\ \ {}^{2}D_{5/2}$ & I & I & I & I & I & I & I & SD & SD &  &  &  &  \\
 &  & I & I & I & I & I & I & S & SD & SD &  &  &  &  \\
30+ & $[Ar]\;3d^{10}\;4s^{2}\;4p^{6}\;4d^{8}\;\ \ {}^{3}F_{4}$ & I & I & I & I & I & I & S & SD & SD &  &  &  &  \\
31+ & $[Ar]\;3d^{10}\;4s^{2}\;4p^{6}\;4d^{7}\;\ \ {}^{4}F_{9/2}$ & I & I & I & I & I & I & S & SD & SD &  &  &  &  \\
32+ & $[Ar]\;3d^{10}\;4s^{2}\;4p^{6}\;4d^{6}\;\ \ {}^{5}D_{4}$ & I & I & I & I & I & I & S & SD & SD &  &  &  &  \\
33+ & $[Ar]\;3d^{10}\;4s^{2}\;4p^{6}\;4d^{5}\;\ \ {}^{4}P_{5/2}$ & I & I & I & I & I & I & S & SD & SD &  &  &  &  \\
34+ & $[Ar]\;3d^{10}\;4s^{2}\;4p^{6}\;4d^{4}\;\ \ {}^{3}P_{0}$ & I & I & I & I & I & I & S & SD & SD &  &  &  &  \\
35+ & $[Ar]\;3d^{10}\;4s^{2}\;4p^{6}\;4d^{3}\;\ \ {}^{4}F_{3/2}$ & I & I & I & I & I & I & S & SD & SD &  &  &  &  \\
36+ & $[Ar]\;3d^{10}\;4s^{2}\;4p^{6}\;4d^{2}\;\ \ {}^{3}F_{2}$ & I & I & I & I & I & I & S & SD & SD &  &  &  &  \\
37+ & $[Ar]\;3d^{10}\;4s^{2}\;4p^{6}\;4d^{1}\;\ \ {}^{2}D_{3/2}$ & I & I & I & I & I & I & S & SD & SD &  &  &  &  \\
 &  & I & I & I & I & I & I & SD & SD & SD &  &  &  &  \\
38+ & $[Ar]\;3d^{10}\;4s^{2}\;4p^{6}\;\ \ {}^{1}S_{0}$ & I & I & I & I & I & I & SD & SD &  &  &  &  &  \\
39+ & $[Ar]\;3d^{10}\;4s^{2}\;4p^{5}\;\ \ {}^{2}P_{3/2}$ & I & I & I & I & I & I & SD & SD &  &  &  &  &  \\
40+ & $[Ar]\;3d^{10}\;4s^{2}\;4p^{4}\;\ \ {}^{3}P_{2}$ & I & I & I & I & I & I & SD & SD &  &  &  &  &  \\
41+ & $[Ar]\;3d^{10}\;4s^{2}\;4p^{3}\;\ \ {}^{2}D_{3/2}$ & I & I & I & I & I & I & SD & SD &  &  &  &  &  \\
42+ & $[Ar]\;3d^{10}\;4s^{2}\;4p^{2}\;\ \ {}^{3}P_{0}$ & I & I & I & I & I & I & SD & SD &  &  &  &  &  \\
43+ & $[Ar]\;3d^{10}\;4s^{2}\;4p^{1}\;\ \ {}^{2}P_{1/2}$ & I & I & I & I & I & I & SD & SD &  &  &  &  &  \\
44+ & $[Ar]\;3d^{10}\;4s^{2}\;\ \ {}^{1}S_{0}$ & I & I & I & I & I & I & SD &  &  &  &  &  &  \\
 &  & I & I & I & I & I & SD & SD &  &  &  &  &  &  \\
45+ & $[Ar]\;3d^{10}\;4s^{1}\;\ \ {}^{2}S_{1/2}$ & I & I & I & I & I & SD & SD &  &  &  &  &  &  \\
 &  & I & I & I & I & SD & SD & SD &  &  &  &  &  &  \\
46+ & $[Ne]\;3s^{2}\;3p^{6}\;3d^{10}\;\ \ {}^{1}S_{0}$ & I & I & I & I & SD & SD &  &  &  &  &  &  &  \\
 &  & I & I & I & S & SD & SD &  &  &  &  &  &  &  \\
47+ & $[Ne]\;3s^{2}\;3p^{6}\;3d^{9}\;\ \ {}^{2}D_{5/2}$ & I & I & I & S & SD & SD &  &  &  &  &  &  &  \\
48+ & $[Ne]\;3s^{2}\;3p^{6}\;3d^{8}\;\ \ {}^{3}F_{4}$ & I & I & I & S & SD & SD &  &  &  &  &  &  &  \\
49+ & $[Ne]\;3s^{2}\;3p^{6}\;3d^{7}\;\ \ {}^{2}D_{5/2}$ & I & I & I & S & SD & SD &  &  &  &  &  &  &  \\
50+ & $[Ne]\;3s^{2}\;3p^{6}\;3d^{6}\;\ \ {}^{5}D_{4}$ & I & I & I & S & SD & SD &  &  &  &  &  &  &  \\
51+ & $[Ne]\;3s^{2}\;3p^{6}\;3d^{5}\;\ \ {}^{4}P_{5/2}$ & I & I & I & S & SD & SD &  &  &  &  &  &  &  \\
52+ & $[Ne]\;3s^{2}\;3p^{6}\;3d^{4}\;\ \ {}^{3}P_{0}$ & I & I & I & S & SD & SD &  &  &  &  &  &  &  \\
53+ & $[Ne]\;3s^{2}\;3p^{6}\;3d^{3}\;\ \ {}^{4}F_{3/2}$ & I & I & I & S & SD & SD &  &  &  &  &  &  &  \\
54+ & $[Ne]\;3s^{2}\;3p^{6}\;3d^{2}\;\ \ {}^{3}F_{2}$ & I & I & I & S & SD & SD &  &  &  &  &  &  &  \\
55+ & $[Ne]\;3s^{2}\;3p^{6}\;3d^{1}\;\ \ {}^{2}D_{3/2}$ & I & I & I & S & SD & SD &  &  &  &  &  &  &  \\
 &  & I & I & I & SD & SD & SD &  &  &  &  &  &  &  \\
56+ & $[Ne]\;3s^{2}\;3p^{6}\;\ \ {}^{1}S_{0}$ & I & I & I & SD & SD &  &  &  &  &  &  &  &  \\
57+ & $[Ne]\;3s^{2}\;3p^{5}\;\ \ {}^{2}P_{3/2}$ & I & I & I & SD & SD &  &  &  &  &  &  &  &  \\
58+ & $[Ne]\;3s^{2}\;3p^{4}\;\ \ {}^{3}P_{2}$ & I & I & I & SD & SD &  &  &  &  &  &  &  &  \\
59+ & $[Ne]\;3s^{2}\;3p^{3}\;\ \ {}^{2}D_{3/2}$ & I & I & I & SD & SD &  &  &  &  &  &  &  &  \\
60+ & $[Ne]\;3s^{2}\;3p^{2}\;\ \ {}^{3}P_{0}$ & I & I & I & SD & SD &  &  &  &  &  &  &  &  \\
61+ & $[Ne]\;3s^{2}\;3p^{1}\;\ \ {}^{2}P_{1/2}$ & I & I & I & SD & SD &  &  &  &  &  &  &  &  \\
 &  & I & S & SD & SD & SD &  &  &  &  &  &  &  &  \\
62+ & $1s^{2}\;2s^{2}\;2p^{6}\;3s^{2}\;\ \ {}^{1}S_{0}$ & I & S & SD & SD &  &  &  &  &  &  &  &  &  \\
63+ & $1s^{2}\;2s^{2}\;2p^{6}\;3s^{1}\;\ \ {}^{2}S_{1/2}$ & I & S & SD & SD &  &  &  &  &  &  &  &  &  \\
 &  & I & SD & SD & SD &  &  &  &  &  &  &  &  &  \\
64+ & $1s^{2}\;2s^{2}\;2p^{6}\;\ \ {}^{1}S_{0}$ & I & SD & SD &  &  &  &  &  &  &  &  &  &  \\
65+ & $1s^{2}\;2s^{2}\;2p^{5}\;\ \ {}^{2}P_{3/2}$ & I & SD & SD &  &  &  &  &  &  &  &  &  &  \\
66+ & $1s^{2}\;2s^{2}\;2p^{4}\;\ \ {}^{3}P_{2}$ & I & SD & SD &  &  &  &  &  &  &  &  &  &  \\
67+ & $1s^{2}\;2s^{2}\;2p^{3}\;\ \ {}^{2}D_{3/2}$ & I & SD & SD &  &  &  &  &  &  &  &  &  &  \\
68+ & $1s^{2}\;2s^{2}\;2p^{2}\;\ \ {}^{3}P_{0}$ & I & SD & SD &  &  &  &  &  &  &  &  &  &  \\
69+ & $1s^{2}\;2s^{2}\;2p^{1}\;\ \ {}^{2}P_{1/2}$ & I & SD & SD &  &  &  &  &  &  &  &  &  &  \\
70+ & $1s^{2}\;2s^{2}\;\ \ {}^{1}S_{0}$ & SD & SD &  &  &  &  &  &  &  &  &  &  &  \\
71+ & $1s^{2}\;2s^{1}\;\ \ {}^{2}S_{1/2}$ & SD & SD &  &  &  &  &  &  &  &  &  &  &  \\
\end{longtable}

\clearpage

\begin{longtable}{@{}l @{\extracolsep{\fill}} l rlr rrl @{}}
\caption{Ionization energies of tungsten, \ce{W^2+} through \ce{W^71+}.}
\label{tab:ip}
Ion & Sequence & \multicolumn{3}{c}{Literature} & Present & \multicolumn{2}{c}{Present recommended values} \\
\cmidrule{3-5}\cmidrule{7-8}
& & Ref.~\cite{Kramida2006} & $\pm$unc. & Ref.~\cite{Beiersdorfer2012} & FAC & GRASP+CI & $\pm$unc. \\
\midrule
\endfirsthead
\multicolumn{8}{l}{Table \ref{tab:ip} (continued)}\\
\midrule
Ion & Sequence & \multicolumn{3}{c}{Literature} & Present & \multicolumn{2}{c}{Present recommended values} \\
\cmidrule{3-5}\cmidrule{7-8}
& & Ref.~\cite{Kramida2006} & $\pm$unc. & Ref.~\cite{Beiersdorfer2012} & FAC & GRASP+CI & $\pm$unc. \\
\midrule
\endhead
\midrule
\endfoot
\bottomrule
\endlastfoot
\ce{W^2+} & Hf & 26.0 & 0.4 & 24.44 & 24.31 & 25.57 & 0.3 \\
\ce{W^3+} & Lu & 38.2 & 0.4 & 37.39 & 36.53 & 37.76 & 0.3 \\
\ce{W^4+} & Yb & 51.6 & 0.3 & 48.60 & 50.13 & 51.28 & 0.3 \\
\ce{W^5+} & Tm & 64.77 & 0.04 & 63.59 & 63.48 & 64.45 & 0.4 \\
\ce{W^6+} & Er & 122.01 & 0.06 & 120.55 & 118.25 & 121.45 & 0.4 \\
\ce{W^7+} & Ho & 141.2 & 1.2 & 139.90 & 139.64 & 139.47 & 0.4 \\
\ce{W^8+} & Dy & 160.2 & 1.2 & 156.13 & 157.59 & 158.89 & 0.3 \\
\ce{W^9+} & Tb & 179.0 & 1.2 & 176.43 & 176.53 & 177.83 & 0.3 \\
\ce{W^10+} & Gd & 208.9 & 1.2 & 206.01 & 205.25 & 208.32 & 0.3 \\
\ce{W^11+} & Eu & 231.6 & 1.2 & 230.31 & 232.14 & 230.21 & 0.3 \\
\ce{W^12+} & Sm & 258.2 & 1.2 & 255.14 & 254.56 & 257.60 & 0.2 \\
\ce{W^13+} & Pm & 290.7 & 1.2 & 291.93 & 287.11 & 289.93 & 0.2 \\
\ce{W^14+} & Nd & 325.3 & 1.5 & 323.59 & 321.81 & 324.41 & 0.2 \\
\ce{W^15+} & Pr & 361.9 & 1.5 & 359.91 & 359.11 & 361.18 & 0.2 \\
\ce{W^16+} & Ce & 387.9 & 1.2 & 385.07 & 386.32 & 387.79 & 0.2 \\
\ce{W^17+} & La & 420.7 & 1.4 & 420.92 & 417.38 & 419.47 & 0.2 \\
\ce{W^18+} & Ba & 462.1 & 1.4 & 458.01 & 458.89 & 460.87 & 0.2 \\
\ce{W^19+} & Cs & 502.6 & 1.4 & 502.17 & 499.21 & 501.17 & 0.2 \\
\ce{W^20+} & Xe & 543.4 & 1.4 & 539.75 & 539.61 & 541.55 & 0.2 \\
\ce{W^21+} & I & 594.5 & 1.5 & 585.92 & 593.08 & 594.97 & 0.2 \\
\ce{W^22+} & Te & 640.7 & 1.5 & 638.63 & 638.15 & 639.92 & 0.2 \\
\ce{W^23+} & Sb & 685.6 & 1.6 & 687.67 & 682.98 & 684.81 & 0.2 \\
\ce{W^24+} & Sn & 734.1 & 1.7 & 729.36 & 731.71 & 733.54 & 0.2 \\
\ce{W^25+} & In & 784.4 & 1.9 & 781.97 & 782.19 & 784.03 & 0.2 \\
\ce{W^26+} & Cd & 833.4 & 1.9 & 825.61 & 830.91 & 831.40 & 0.2 \\
\ce{W^27+} & Ag & 881.4 & 1.6 & 879.19 & 879.12 & 879.61 & 0.2 \\
\ce{W^28+} & Pd & 1132.2 & 1.4 & 1128.27 & 1128.13 & 1129.31 & 0.2 \\
\ce{W^29+} & Rh & 1179.9 & 1.4 & 1180.89 & 1175.32 & 1176.64 & 0.2 \\
\ce{W^30+} & Ru & 1230.4 & 1.4 & 1225.59 & 1225.86 & 1227.10 & 0.2 \\
\ce{W^31+} & Tc & 1283.4 & 1.4 & 1280.78 & 1279.44 & 1280.49 & 0.2 \\
\ce{W^32+} & Mo & 1335.1 & 1.4 & 1326.64 & 1330.99 & 1332.07 & 0.2 \\
\ce{W^33+} & Nb & 1386.7 & 1.4 & 1381.82 & 1382.52 & 1383.62 & 0.2 \\
\ce{W^34+} & Zr & 1459.9 & 1.5 & 1454.82 & 1454.90 & 1456.07 & 0.2 \\
\ce{W^35+} & Y & 1512.4 & 1.5 & 1511.73 & 1507.59 & 1508.82 & 0.2 \\
\ce{W^36+} & Sr & 1569.1 & 1.5 & 1559.65 & 1564.66 & 1565.74 & 0.2 \\
\ce{W^37+} & Rb & 1621.7 & 1.5 & 1617.82 & 1617.79 & 1618.67 & 0.2 \\
\ce{W^38+} & Kr & 1829.8 & 1.9 & 1829.18 & 1828.93 & 1830.66 & 0.2 \\
\ce{W^39+} & Br & 1883.0 & 2.0 & 1887.10 & 1881.81 & 1883.71 & 0.2 \\
\ce{W^40+} & Se & 1940.6 & 2.0 & 1935.69 & 1938.71 & 1941.73 & 0.2 \\
\ce{W^41+} & As & 1994.8 & 2.0 & 1994.53 & 1994.37 & 1995.84 & 0.2 \\
\ce{W^42+} & Ge & 2149.2 & 2.1 & 2144.78 & 2145.23 & 2146.18 & 0.2 \\
\ce{W^43+} & Ga & 2210.0 & 1.5 & 2206.13 & 2206.32 & 2206.67 & 0.2 \\
\ce{W^44+} & Zn & 2354.5 & 1.4 & 2351.89 & 2351.82 & 2354.58 & 0.2 \\
\ce{W^45+} & Cu & 2414.1 & 0.4 & 2413.51 & 2413.42 & 2414.63 & 0.3 \\
\ce{W^46+} & Ni & 4057 & 3 & 4051.67 & 4051.32 & 4053.03 & 0.4 \\
\ce{W^47+} & Co & 4180 & 4 & 4187.70 & 4173.08 & 4175.00 & 0.4 \\
\ce{W^48+} & Fe & 4309 & 4 & 4302.64 & 4303.39 & 4305.01 & 0.3 \\
\ce{W^49+} & Mn & 4446 & 4 & 4445.11 & 4442.30 & 4443.24 & 0.3 \\
\ce{W^50+} & Cr & 4578 & 4 & 4562.79 & 4575.37 & 4575.91 & 0.3 \\
\ce{W^51+} & V & 4709 & 4 & 4705.14 & 4705.82 & 4706.42 & 0.3 \\
\ce{W^52+} & Ti & 4927 & 4 & 4919.64 & 4919.72 & 4920.54 & 0.2 \\
\ce{W^53+} & Sc & 5063 & 4 & 5067.35 & 5055.47 & 5056.10 & 0.2 \\
\ce{W^54+} & Ca & 5209 & 4 & 5190.91 & 5204.05 & 5204.06 & 0.2 \\
\ce{W^55+} & K & 5348 & 4 & 5341.79 & 5341.66 & 5341.16 & 0.2 \\
\ce{W^56+} & Ar & 5719 & 5 & 5726.70 & 5725.80 & 5729.10 & 0.2 \\
\ce{W^57+} & Cl & 5840 & 5 & 5858.13 & 5846.25 & 5849.69 & 0.2 \\
\ce{W^58+} & S & 5970 & 5 & 5968.91 & 5978.45 & 5981.23 & 0.2 \\
\ce{W^59+} & P & 6093 & 5 & 6102.18 & 6101.28 & 6103.87 & 0.2 \\
\ce{W^60+} & Si & 6596 & 7 & 6584.47 & 6584.57 & 6586.12 & 0.2 \\
\ce{W^61+} & Al & 6735 & 7 & 6725.60 & 6724.91 & 6725.87 & 0.3 \\
\ce{W^62+} & Mg & 7000 & 7 & 6997.72 & 6997.54 & 7001.66 & 0.3 \\
\ce{W^63+} & Na & 7130 & 7 & 7131.33 & 7130.98 & 7130.93 & 0.14 \\
\ce{W^64+} & Ne & 15566 & 11 & 15603.58 & 15600.92 & 15604.60 & 0.12 \\
\ce{W^65+} & F & 15896 & 11 & 15965.53 & 15930.67 & 15934.42 & 0.12 \\
\ce{W^66+} & O & 16252 & 11 & 16266.94 & 16295.54 & 16298.19 & 0.10 \\
\ce{W^67+} & N & 16588 & 11 & 16635.31 & 16632.49 & 16635.00 & 0.10 \\
\ce{W^68+} & C & 18476 & 4 & 18417.59 & 18417.15 & 18418.85 & 0.10 \\
\ce{W^69+} & B & 18872 & 3 & 18829.80 & 18828.94 & 18821.05 & 0.10 \\
\ce{W^70+} & Be & 19362 & 3 & 19348.69 & 19348.72 & 19358.63 & 0.10 \\
\ce{W^71+} & Li & 19691 & 3 & 19686.94 & 19686.40 & 19686.68 & 0.07 \\
\end{longtable}

\end{document}